\documentclass[iop,apjl]{emulateapj}
\journalinfo{The Astrophysical Journal Letters} 
\slugcomment{accepted for publication}

\shorttitle{Turbulent Interchange Reconnection}
\shortauthors{Rappazzo et al.}

\begin{document}

\title{Interchange reconnection in a turbulent  Corona}

\author{A. F. Rappazzo$^{1,\dagger}$, W. H. Matthaeus$^1$, D. Ruffolo$^{2,3}$, 
S. Servidio$^4$, and M. Velli$^5$}
\affil{$^1$Bartol Research Institute, Department of Physics and Astronomy, 
University of Delaware, Delaware 19716, USA\\
$^2$Department of Physics, Faculty of Science, Mahidol University, Bangkok 10400, Thailand\\
$^3$Thailand Center of Excellence in Physics, CHE, Ministry of Education, Bangkok 10400, Thailand\\
$^4$Dipartimento di Fisica, Universit\`a della Calabria, I-87036 Cosenza, Italy\\
$^5$Jet Propulsion Laboratory, California Institute of Technology, Pasadena, CA 91109, USA}
\email{$^{\dagger}$rappazzo@udel.edu}

\begin{abstract}
Magnetic reconnection at the interface between coronal holes
and loops, so-called interchange reconnection, can release
the hotter, denser plasma from magnetically confined 
regions into the heliosphere, contributing to the formation of
the highly variable slow solar wind. 
The interchange process is  often thought to develop  
at the apex of streamers or pseudo-streamers, near 
$Y$ and $X$-type neutral points, but 
slow streams with loop composition have been 
recently observed along fanlike open field lines adjacent to
closed regions, far from the apex.
However, coronal heating models, with magnetic field lines 
shuffled by convective motions, show that reconnection can 
occur continuously in unipolar magnetic field regions with no 
neutral points: photospheric motions induce a 
magnetohydrodynamic turbulent cascade in the coronal field
that creates the \emph{necessary}
small scales, where a \emph{sheared} magnetic field component orthogonal
to the strong axial field is created locally and can reconnect.
We propose that a similar mechanism operates near and around
boundaries between open and closed regions inducing a continual 
stochastic rearrangement of connectivity.
We examine a reduced magnetohydrodynamic 
model of a simplified interface region 
between open and closed corona 
threaded by a strong unipolar magnetic field.
This boundary is not stationary, becomes fractal, and field lines 
change connectivity continuously, becoming alternatively open and 
closed.  
This  model suggests that 
slow wind may originate everywhere
along loop-coronal hole boundary regions, 
and can account naturally and simply
for outflows at and adjacent to such boundaries
and for the observed diffusion of  slow wind around the 
heliospheric current sheet.
\end{abstract}


\section{Introduction}

\begin{figure}
\begin{centering}
\includegraphics[scale=.35]{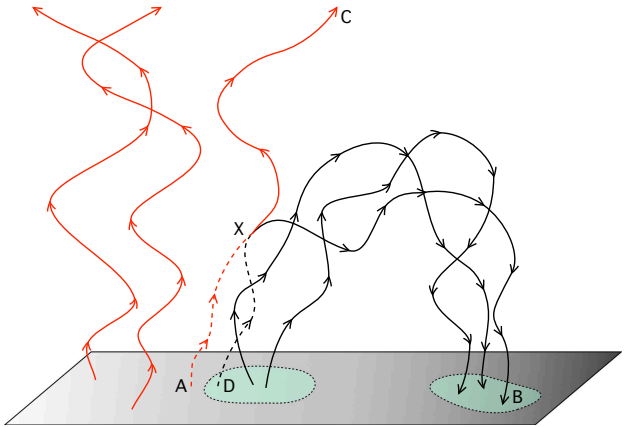}
\caption{Cartoon of interchange reconnection in a turbulent Corona.
Reconnection takes place in point X, along the boundary between open
and closed corona, with closed field line A-X-B opening into A-X-C, 
while the field line traced from point B closes at a different location, point D, 
forming the closed field line D-X-B. Swinging field lines depict
magnetic fluctuations (pictorially exaggerated).
\label{fig:fig1}}
\end{centering}
\end{figure}

A topic of recent interest is magnetic reconnection
between open and closed field lines at the interface between 
coronal holes and loops, dubbed ``interchange reconnection'' 
(Figure~\ref{fig:fig1}, IR hereafter).
This mechanism can contribute mass, heat, and momentum
to the solar wind, with numerous 
heliospheric implications
\citep{fzs99, FiskSchwadron01, cgk02, de03, ant07,
owe08, edm09, lin11, tit11, mas12}.

The solar wind may be classified as ``fast'',
when velocities $v$ exceed, say,  $600\ \textrm{km}\, \textrm{s}^{-1}$, and ``slow'', 
when $v < 500\ \textrm{km}\, \textrm{s}^{-1}$. The steadier fast wind originates in 
polar coronal holes (dark X-ray regions) and similar 
open-field regions closer to the equator, propagating radially into 
interplanetary space \citep{zir77}. The more spatially and temporally 
intermittent slow wind originates in and around the coronal 
streamer belt \citep{wang94}, where a significant 
population of closed-field structures is found.
The heliospheric current sheet (HCS) is always 
embedded within slow wind, which 
surrounds it in a region 
spanning about $30^{\circ}$ in latitude near 
solar minimum conditions \citep{gos81, bor81, win94}.

Fast and slow wind differ 
in their plasma composition. 
Generally the fast wind composition is similar
to that of the photosphere, while slow wind 
composition 
is similar to that of coronal loops, with comparable
abundances ratios of low to high first ionization potential (FIP) elements
and ions with different charge states (e.g.,  O$^{7+}$/O$^{6+}$)
\citep{gei95, zfgs02, fw03}.
IR allows field-line 
connectivity to change from closed to open, thus 
releasing coronal loops plasma
into the heliosphere. 
This has been suggested as a primary mechanism
for the formation of the slow wind \citep{wang98, ant11}.

Recent observations motivate 
to advance  our understanding of the physical 
processes at the root of IR. 
Outflows along open fanlike field lines, 
at the edges of active (closed) regions, 
have been observed by \emph{Hinode}
EUV imaging spectrometer (EIS) measurements \citep{sak07}.
\citet{bw11} found that the composition of these 
outflows is that of coronal loops and established a link with
slow wind detected in situ by the Solar Wind Ion Composition 
Spectrometer on board the Advanced Composition 
Explorer (ACE).
Outflows in the streamer belt region are observed by the
LASCO-C2 white light coronagraph \citep{wgrs12} 
and STEREO imagers \citep{hdr12},
suggesting that mixing and dynamics contribute to the 
average observed configuration. 
 
In most prior work, IR has been 
thought to occur only in special  topological locations, 
at the apex of streamers and pseudo-streamers 
corresponding to $Y$ or $X$-points \citep{wgrs12},
where field lines of \emph{opposite polarity} can reconnect in 
a neutral point with $\mathbf{B}=0$.

\cite{wang98} proposed that convective field line shuffling
can trigger IR around the cusp region, resulting in
the outward propagation of density enhanced blobs.
But the physical mechanism leading
to and allowing IR remains undetermined.
In fact, the small value of resistivity in the solar corona implies
that magnetic field lines are frozen in the plasma except
where  very \emph{small scales} are present, i.e., 
strong currents with an apt local magnetic field topology
for magnetic reconnection to occur. 

Numerical simulations \citep{evpp96, dg97, rved07} 
of Parker model for coronal heating \citep{park72, park88}
have shown that the continuous shuffling of magnetic field 
lines' footpoints by photospheric convective motions induces a
magnetohydrodynamic (MHD) turbulent cascade in the unipolar closed
coronal field with \emph{no null points}.
This cascade transfers energy from the large to 
the small scales, driving field-aligned current sheets that
are continuously formed and dissipated, 
where the magnetic field component orthogonal to the strong 
axial field is sheared, i.e., its field lines are locally oppositely 
directed, and can reconnect (\emph{nanoflares}).

Therefore the \emph{unipolar closed field lines} of coronal loops 
\emph{continuously change connectivity} due to this
dynamical activity. 
Furthermore, in view of ubiquitous 
presence of magnetic fluctuations in this scenario, 
at each instant of time the magnetic field lines admit a random character
due to field line \emph{random-walk} \citep[FLRW:][]{JokipiiParker68, Jokipii69, mbr07}.
In this environment, magnetic connectivity is very complex
and changing.

Here we suggest that similar dynamics take place everywhere
at the boundary between open and closed regions where 
turbulent IR can occur stochastically 
(Figures~\ref{fig:fig1}-\ref{fig:fig2}), 
naturally accounting for the observed 
flows along and around these boundaries, including those at
adjacent active region edges observed by \cite{sak07},
that cannot be explained by IR at the streamer apex.

In this paper we investigate the dynamics of IR 
at the interface between open and closed corona, with
photospheric convective motions shuffling the magnetic 
field lines' footpoints. For a simple first demonstration,
we apply photospheric motions only to the (originally) 
closed region, so that no waves or turbulent dynamics
are excited directly by photospheric motions along 
the originally open field lines.

\section{Model and Governing Equations} \label{sec:md}

\begin{figure}
\begin{centering}
\includegraphics[scale=.45]{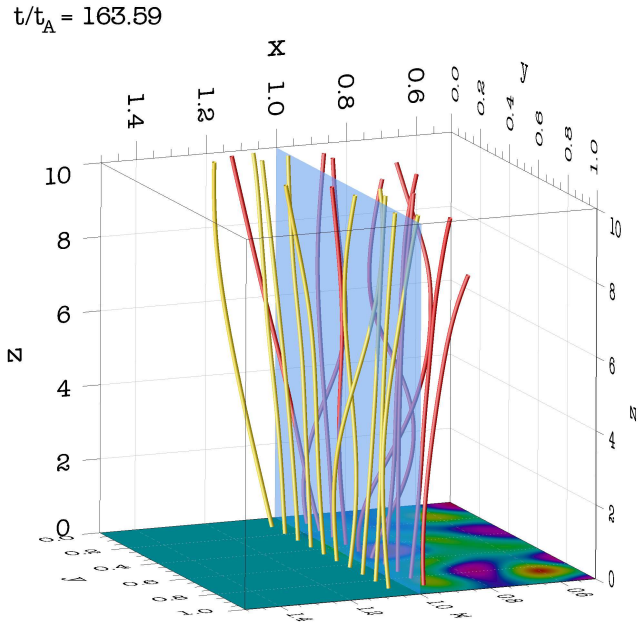}
\caption{Snapshot of the simulated interface between closed 
(straightened out loop)  and open regions of the solar Corona.
Field lines are line-tied at the photospheric plane $z=0$,
where convection-mimicking motions (shown in contours)
are applied at $x<1$ and vanish for $x>1$.
Closed field lines return to the plate $z=10$ for $x < 1$
where they are line-tied to a motionless photosphere,
while for $x > 1$ an open boundary is realized.
The plane $x=1$ is the original boundary magnetic surface 
between open and closed regions at $t=0$.
\label{fig:fig2}}
\end{centering}
\end{figure}

We model the interface region in \emph{Cartesian geometry} (Figure~\ref{fig:fig2}),
with a straightened loop juxtaposed with an open-field 
region. Curvature effects are neglected.
The computational box spans $0< x <2$, $ 0< y<1$, and $0<z<10$ 
(lengths are normalized by $\ell^{\ast} = 60\, \textrm{Mm}$, e.g., $z$ spans 
$600\, \textrm{Mm}$).
The plane $z=0$ represents the photosphere where  both open and
closed field lines are line-tied. On the $z=0$  plane, 
in the $x < 1$ region,  convective super-granular
motions are modeled by imposing 
a large-scale velocity pattern with all modes with wave-numbers
between 3 and 4 excited with random amplitudes
and normalized to have an r.m.s.\ value $\sim 0.5\ \textrm{km}\, \textrm{s}^{-1}$,
that in physical space correspond to distorted vortical
streamlines with length $\ell_c \sim 15\, \textrm{Mm}$.
This model of footpoint motion is 
similar to that of \cite{rved08}, and is illustrated 
in the contours at the bottom of Figure~\ref{fig:fig2}.
On the remaining region of the $z=0$ plane, 
where  $x > 1$,  the velocity vanishes.
The upper plane at $z=10$, in the region 
$x<1$, represents  
the photospheric plate where closed loop field lines 
return to, and are 
line-tied to a \emph{motionless} photosphere.
On the section of the $z=10$ plane 
having $x>1$,  an \emph{open
boundary} is realized, imposing non-reflecting
boundary conditions \citep{thom87, thom90, vtb89}, i.e.,
wave-like signals are allowed to propagate out toward $z>10$ with no reflection 
toward $z<10$ \citep[e.g.,][]{rved05}. Along $x$ and $y$ periodic
boundary conditions are used.

The above system is threaded by 
a strong and uniform unipolar magnetic field 
$\mathbf{B_0} = B_0\, \mathbf{\hat{e}}_z$
along $z$. The
field lines traced from the bottom photospheric plane $z=0$
are considered to be either \emph{closed} when they map
to the top $z=10$ plate with $x < 1$, or \emph{open}
for $x > 1$. 
Because of a large assumed conductivity
and line-tying, 
a field line must undergo 
magnetic reconnection to change connectivity.
Field lines traced from the plate $z=10$ with $x<1$ map
the actual closed region in the $z=0$ plane. 
Likewise those traced from $z=10$ with $x>1$
map the open region back to the $z=0$ plane. 

As in previous work the dynamics are integrated with 
the (nondimensional) equations of reduced 
magnetohydrodynamics (MHD) \citep{kp74,str76,mon82},
well suited for a plasma embedded in a strong 
axial magnetic field:{\setlength\arraycolsep{-10pt}
\small{\begin{eqnarray}
&& \partial_t \mathbf{u}  + 
 \mathbf{u} \cdot \nabla  \mathbf{u} = 
- \nabla P + \mathbf{b} \cdot \nabla  \mathbf{b}
+ c_{_{\!A}} \partial_z \mathbf{b}
+ \frac{ (-1)^{n+1} }{Re_n} \nabla^{2n} \mathbf{u}, 
\label{eq:eq1} \\
&&\partial_t \mathbf{b}  + 
\mathbf{u} \cdot \nabla \mathbf{b} = 
\mathbf{b} \cdot \nabla \mathbf{u} 
+ c_{_{\!A}} \partial_z \mathbf{u}
+ \frac{ (-1)^{n+1} }{Re_n} \nabla^{2n} \mathbf{b}, 
\label{eq:eq2}
\end{eqnarray}
}}with $\nabla \cdot \mathbf{u} =  \nabla \cdot \mathbf{b} = 0$.  
Here, gradient and Laplacian operators have only transverse  
($x$-$y$) components as do velocity and magnetic field vectors 
($u_z = b_z = 0$), while $P$ is the
total (plasma plus magnetic) pressure, $c_{_{\!A}}$ 
is the Alfv\'en velocity of the axial field ($B_0/\sqrt{4 \pi \rho_0}$), the plasma 
is assumed to have uniform density $\rho_0$.
In the simulation presented here $c_{_{\!A}}=200$
(velocities are normalized by $u^{\ast} = 0.5\ \textrm{km}\, \textrm{s}^{-1}$),
and the numerical grid has $512 \times 256 \times 120$ points
to achieve the long duration of $\sim 1,800$ Alfv\'en crossing times 
$\tau_A = L_z/c_{_{\!A}}$ ($L_z = 10$ is the loop length),
corresponding to $\sim 180$ nonlinear times, necessary to acquire significant statistics.
We use hyperdiffusion with $n=4$ and $R_4 = 5\, \times 10^{16}$ that eliminates
diffusion at the large scales, a critical feature in this kind of simulations that
otherwise reach a diffusive regime 
\cite[see, also for a more detailed description of the numerical code,][]{rved08}.

\section{Results} \label{sec:ns}

Initially photospheric motions induce magnetic 
fluctuations in the closed
regions that grow linearly in time. 
A nonlinear, turbulent stage is attained 
around time $\tau_{_{\!NL}} \sim 10\, \tau_A$
\citep[for details see][]{rved08}. 
The dynamics subsequently
leads to many field-aligned current 
sheets where magnetic reconnection occurs.
These structures are highly dynamic, 
crossing a transverse 
correlation length (here, 
the super-granulation scale $\ell_c$) 
in approximately 
a nonlinear time-scale $\tau_{_{\!NL}}$.
In this dynamic sea of current sheets,
roughly half of those 
forming within a correlation length 
from the open-closed boundary will
encounter it, thus changing its location and inducing 
changes in field-line connectivity.

\begin{figure}
\begin{centering}
\includegraphics[scale=.44]{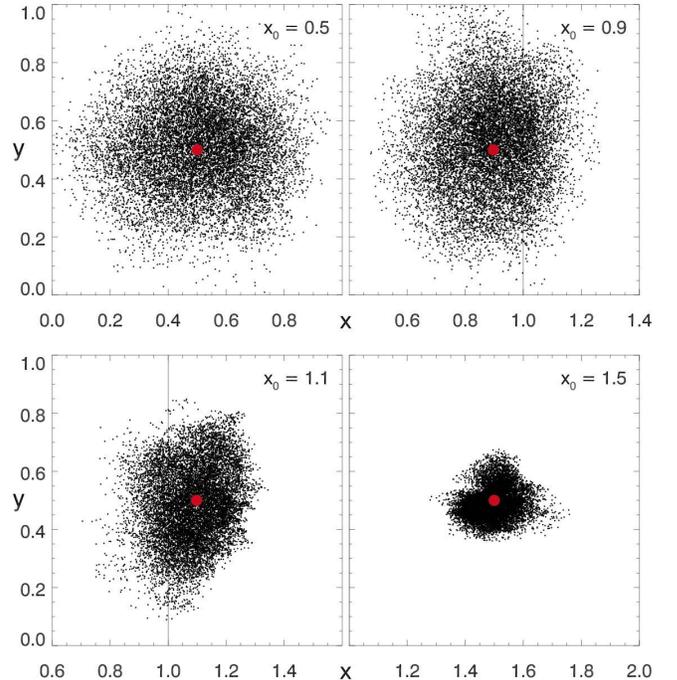}
\caption{Field lines' footpoints in the plane $z=10$.
Field lines are traced from points with $x=x_0$ in the plane $z=0$. 
Their footpoints' orthogonal displacement
relative to their origination point (central circle) is shown for four selected values
of $x_0$. The boundary between open and closed regions is at $x=1$.
\label{fig:fig3}}
\end{centering}
\end{figure}

Figure~\ref{fig:fig2} shows field lines at $t \sim 163.59\, \tau_A$
originating at $z=0$ 
near the initial open-closed boundary 
at $t=0$ ($x=1$), but which 
cross the boundary prior to arrival at $z=10$. 
This opening and closing of 
field lines is caused by 
the highly dynamic current sheets and 
reconnection described above.
This is bursty and stochastic, 
and increasingly so with higher Reynolds 
numbers \citep{sms09}.

To quantitatively understand  the impact of this kind of IR
we analyze the statistical properties of field lines.
We ask what is the fraction of time spent in open/closed 
regions, or the probability
for a field line traced from a point 
$x=x_0$ from the boundary 
in the photospheric plane $z=0$ to be closed
or open when it arrives at the top plane 
$z=10$. The latter condition corresponds to 
$x<1$ or $x>1$ respectively. 

To address this, 
we trace field lines from points in the plane 
$z=0$ with $x=x_0$ fixed. 
They are traced at $320$ different times, 
separated by $\Delta t = 5\, \tau_A$, corresponding to 
approximately half a nonlinear time. To increase statistics 
the field lines are computed in 40 equally spaced points 
along $y$ (along this direction points are statistically equivalent).
Figure~\ref{fig:fig3} shows the footpoints'  displacement 
in the $x$-$y$ plane for 12,800 field lines
in the plane $z=10$. Four cases correspond to 
four selected initial distances $x_0$ from the boundary.
The field line tracing code \citep{dal12} employs a fifth order 
Runge-Kutta with adaptive step-size, and second order interpolation.

Field lines traced from the middle of the originally  closed and open
regions, i.e., $x_0 = 0.5$ and $1.5$, exhibit an isotropic distribution
of footpoints of different extension because in the closed region
magnetic fluctuations are stronger. No waves are injected from 
the region $z=0$ with $x>1$, 
but magnetic field fluctuations ``leak'' from the closed region,
where magnetic forces push the magnetic islands against 
each other and these forces are unbalanced by weaker fields
in the open region, 
with the energy density reducing across the original boundary 
at $x=1$. This drop in turbulence intensity
along $x$ is the cause of anisotropy for the footpoints distributions
for $x_0=0.9$ and $1.1$ that are narrower along $x$.

\begin{figure}
\begin{centering}
\includegraphics[scale=.4]{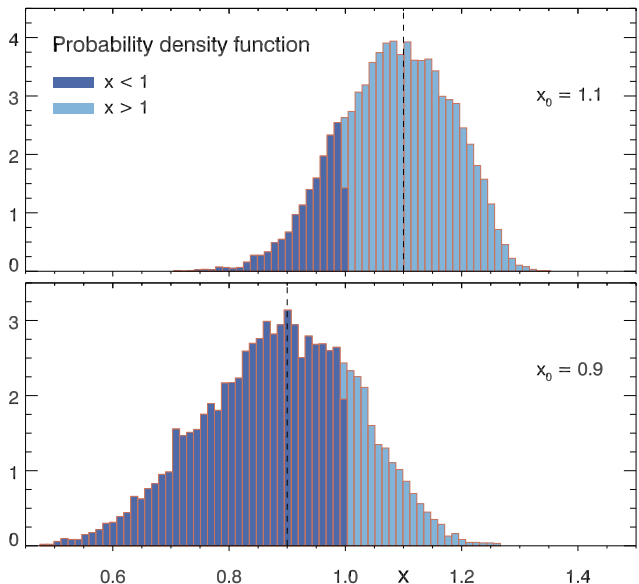}
\caption{Probability density functions of the footpoints shown in Figure~\ref{fig:fig3} for $x_0=0.9$
and $x_0 = 1.1$, across the boundary between open and closed regions at
$x=1$.
\label{fig:fig4}}
\end{centering}
\end{figure}

Field lines traced from $x_0=0.5$ and $1.5$, more distant 
from the initial open-closed boundary, never 
change connectivity from open to closed (or vice-versa).
But those traced from $x_0=0.9$ and $1.1$ do, and their p.d.f.s
are shown in Figure~\ref{fig:fig4}. They have a probability of $\sim 20\%$
to change connectivity from their initial type at time $t=0$.
From the dynamical point of view the probability
can be seen as the fraction of time the field line is closed or open.

Given the complex character of the magnetic
field $\mathbf{b}$, which becomes broad band in space, 
while also evolving in time, 
it is natural to try to describe the spreading 
(and changes of connectivity) as a \emph{diffusive} process. 
In fact there are two related diffusive processes at work.
First, in Figure~\ref{fig:fig2} we see that at a fixed instant of time, 
field lines wander randomly due to fluctuations.  
Second, for the spreading found 
in the above numerical experiment shown 
in Figures~\ref{fig:fig3} and \ref{fig:fig4}, the time dependence of the 
turbulence contributes an additional randomizing effect.
The diffusive nature of the field line spreading 
is verified by the computation (not shown)
of the mean square displacement $\langle \Delta x^2 \rangle$, 
which  increases linearly with height in the manner of diffusion:
\begin{equation} \label{eq:diff}
\langle \Delta x^2 \rangle = 2 D z.
\end{equation}
We found this linear scaling for all values of $x_0$, with diffusion
coefficients, e.g., 
$D = 93.6$, $56.82$ and $14.46\, \textrm{km}$ 
for $x_0 = 0.5$, 1, and $1.5$ respectively
(normalized to $\ell^{\ast} = 60\, \textrm{Mm}$).

This type of randomization of magnetic field lines
is not often associated with coronal fields 
(usually modeled as smooth and laminar),
where field lines may be line-tied
at both ends. This diffusive 
rearrangement of connectivity 
\emph{requires} magnetic reconnection to occur.
A quantitative theory for this space-time diffusion of field lines 
appears to be tractable but lengthy, 
and we will address it in a subsequent paper
(Ruffolo et al., in preparation). 

Two features of the diffusion 
theory are pertinent at present. 
First, the expected $\langle \Delta x^2 \rangle$ at height $z$
due to single time randomization -- the 
FLRW effect -- and the expected 
additional mean square displacement 
at height $z$ due to the time dependent 
changing of field lines,
are of the same order. 
Second, there are two broad classes of FLRW
diffusion coefficients, say 
$D_{ql} \sim \lambda_z  (b / B_0)^2$, the quasi-linear result
\citep{JokipiiParker68}, and $D_{Bohm} \sim \lambda_{\perp} b/B_0$
\citep{ghi11}.
Here $\lambda_z \sim 5$ and $\lambda_\perp \sim 1/4$ are suitable 
coherence scales in directions parallel and 
perpendicular to ${\bf B}_0$, with $b/B_0 \sim 2\%$ in the closed region.
Note that
the present case of reduced MHD
requires the ordering 
$b \lambda_z /( B_0 \lambda_{\perp} ) \sim \tau_{\parallel}/\tau_{\perp} \sim 1$, 
from which $D_{Bohm} \sim D_{ql}$.
Consequently
we anticipate that the observed 
diffusive spreading (Figures~\ref{fig:fig2}-\ref{fig:fig4})
is characterized by a diffusion coefficient
on the order of the quasi-linear result.

Also important is 
the evolving structure 
of the boundary between open and closed regions.
At time $t=0$ the boundary is simply the plane $x=1$ (Figure~\ref{fig:fig2}),
but at later times the magnetic 
surface has to be computed.

\begin{figure}
\begin{centering}
\includegraphics[scale=.45]{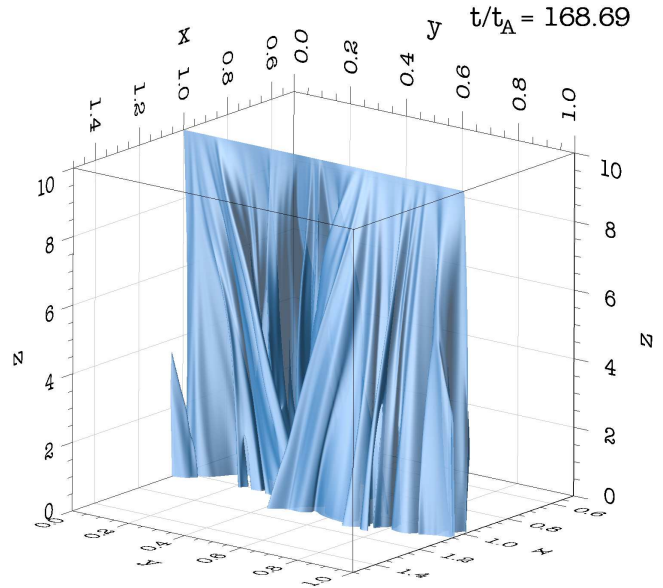}
\caption{Open-closed regions boundary magnetic surface.
\label{fig:fig5}}
\end{centering}
\end{figure}

In reduced MHD,  with uniform axial field 
$\mathbf{B}_0 = B_0 \mathbf{\hat{e}}_z$ and transverse 
fluctuations $\mathbf{b}$, 
the magnetic surface coordinate $\psi$ 
obeys
the magnetic differential equation
\begin{equation} \label{eq:ms}
\partial_z \psi = - \frac{1}{B_0}\ \mathbf{b} \cdot \nabla \psi,
\label{eq:passive}
\end{equation}
where the right side involves only the components
of $\nabla$ transverse to ${\bf B}_0$,
with initial condition $\psi(x,y,z=10) = sin(\pi x)$.
The numerical code employs a third order Runge-Kutta,
quadratic interpolation, and adaptive step-size. 

Like a passive scalar,  
solutions to equation~(\ref{eq:ms})
can acquire a complex structure \citep{mgpb95}.
The boundary magnetic surface $\psi =0$ separates
the two topologically different regions, and in 
Figure~\ref{fig:fig5} it is shown at time $t =168.69\, \tau_A$. 
Its structure is fractal, appears like a pleated drape with many
intricate folds, but for a continuous field $\mathbf{b}$ it does not tear
no matter how folded it is, although numerically a small 
diffusivity  removes the smallest-scale folds and minor 
tearing occurs.
The boundary magnetic surface evolves in time
and on the average its map on the plane $z=0$ has
an excursion in $x$ given by twice $\langle \Delta x^2 \rangle^{1/2} = (2 D L_z)^{1/2}$ 
(equation~(\ref{eq:diff})), where $L_z$ is the loop length.

\section{Conclusions and Discussion} \label{sec:con}

Previous studies
have shown that IR is required, e.g., to explain the quasi-rigid rotation
of coronal holes in presence of the underlying photospheric 
differential rotation \citep{whs96, lio05, lio06},
but this approach \citep[see also][]{sd03, ws04} assumes
a quasi-steady coronal response to photospheric evolution.
In the prevailing view that coronal interchange occurs 
at the apex of streamers and pseudo-streamers in 
correspondence of $Y$ or $X$-points \citep{wgrs12},
all previous simulations and modeling
have used \emph{smooth large-scale fields}
that contain neutral points (${\bf B} = 0$).

\cite{wang98} anticipated that field line footpoints
shuffling may promote IR at the boundary between open
and closed regions.
The present model provides a specific mechanism for this to 
occur, modeling IR as component reconnection
that may occur all along the magnetic 
interface between coronal hole and loop 
threaded by a unipolar field that contain no true neutral points,
and extends the range of occurrence of IR proposed by \cite{wang98}.
These reconnection sites are well known in the context of 
nanoflare models, but their role in interchange has not 
been previously emphasized. 
Here we simulate only a small volume of  the open-closed 
region interface to employ a higher spatial resolution.
This allows the development of MHD turbulence,
with its associated \emph{magnetic fluctuations} 
in the coronal field (Figures~\ref{fig:fig1} and \ref{fig:fig2}),
naturally induced by photospheric motions shuffling
the field lines' footpoints.

Turbulent IR renders the boundary between 
open and closed regions dynamic (Figures~\ref{fig:fig3}-\ref{fig:fig5}).
The boundary fluctuates continuously with an average 
displacement of the order of the 
super-granulation scale $\ell_c \sim 15\, \textrm{Mm}$.
IR can then inject loop plasma along the boundary and 
in the fanlike
regions adjacent to closed regions,
where slow streams with loop composition have 
been recently observed \citep{sak07,bw11},
providing an alternate mechanism to account 
for the plasma composition at the edges of
active regions \citep{cvve07}, and additional
momentum, mass and energy for the streams
originating from there \citep{wang94}.

In a realistic geometry
the field lines originating from this small fanlike
regions expand super-radially in the heliosphere 
and map in an extended region around the HCS.
Thus flows due to turbulent IR naturally diffuse 
away from the HCS, 
overcoming the restriction
of the smooth field model proposed by \cite{ant11}
that admit diffusion only for streams originating
from narrow open flux channels connecting two
coronal holes.

In summary, when field lines' footpoints are shuffled by photospheric 
motions, component magnetic reconnection is expected 
to occur in unipolar loop and open field regions 
and near the boundaries between them. 
This stochastic IR is likely to operate 
all along these boundaries and adjacent regions, where
closed and open field lines
can then \emph{continuously change connectivity}.
On this basis we suggest that plasma and energy transport along these 
magnetic field lines may be an important factor in generating 
the slow wind, and in broadening the regions in which 
compositional and other properties are mixed in the 
solar wind.

In the future we plan to extend this work to more realistic
reduced and full MHD models that include curvature and
expansion effects and alternate boundary conditions,
allowing us to determine the relative importance of apex neutral point
IR and stochastic component IR.

\acknowledgments

This research supported in part by NASA Heliophysics Theory program NNX11AJ44G, 
NSF Solar Terrestrial and SHINE programs AGS-1063439
\& AGS-1156094,  NASA MMS, Solar probe Plus Projects, the Thailand Research Fund
and by EU project ``Turboplasmas''.
Work carried out in part at the Jet Propulsion Laboratory under a contract with NASA.
Simulations performed through the NASA Advanced 
Supercomputing SMD awards 11-2331 and 12-3188.

\end{document}